\documentclass[journal]{IEEEtran}
\usepackage{mathrsfs}
\usepackage{bbm}
\usepackage{amssymb}
\usepackage{bbding}
\usepackage{array}
\usepackage[mathcal]{euscript}
\usepackage{booktabs,caption}
\usepackage[flushleft]{threeparttable}
\usepackage{psfrag,calc,url,bm}
\usepackage{cite}
\usepackage{graphicx}
\usepackage{hyperref}
\usepackage{url}
\usepackage{stfloats}
\usepackage{amsmath}
\usepackage{float}
\usepackage{algorithmic}
\usepackage[ruled,linesnumbered,vlined]{algorithm2e}
\usepackage{color}
\usepackage{boxedminipage}
\usepackage{amsthm}
\usepackage{multirow}

\usepackage{setspace}

\usepackage{soul}
\usepackage{epstopdf}

\usepackage{caption}
\usepackage{booktabs,subcaption,amsfonts,dcolumn}

\allowdisplaybreaks[3]

\begin{document}

\title{Graph Neural Networks for Wireless Networks: \\Graph Representation, Architecture and Evaluation}

\author{Yang Lu, Yuhang Li, Ruichen Zhang, Wei Chen, Bo Ai, and Dusit Niyato
\thanks{This work was supported  in part by the Beijing Nova Program under Grant Z211100002121139, and in part by the National Natural Science Foundation of China under Grant 62101025,  62221001 and 62122012. ({\it Corresponding author: Yuhang Li.}) }
\thanks{Yang Lu and Yuhang Li are with the School of Computer and Technology, and also with the Collaborative Innovation Center of Railway Traffic Safety, Beijing Jiaotong University, Beijing 100044, China (e-mail: yanglu@bjtu.edu.cn, 24110137@bjtu.edu.cn).}
\thanks{Ruichen Zhang and Dusit Niyato are with the College of Computing and Data Science, Nanyang Technological University, Singapore 639798 (e-mail: ruichen.zhang@ntu.edu.sg, dniyato@ntu.edu.sg).}
\thanks{Wei Chen and Bo Ai are with the School of Electronic and Information Engineering, Beijing Jiaotong University, Beijing 100044, China (e-mail: weich@bjtu.edu.cn, boai@bjtu.edu.cn).}
}

\maketitle

\begin{abstract}
Graph neural networks (GNNs) have been regarded as the basic model to facilitate deep learning (DL) to revolutionize resource allocation in wireless networks. GNN-based models are shown to be able to learn the structural information about graphs representing the wireless networks to adapt to the time-varying channel state information and dynamics of network topology. This article aims to provide a comprehensive overview of applying GNNs to optimize wireless networks via answering three fundamental questions, i.e., how to input the system parameters of wireless networks into  GNNs, how to improve the expressive performance of GNNs, and how to evaluate GNNs. Particularly, two graph representations are given to transform  wireless network parameters into graph-structured data. Then, we focus on the architecture design of the GNN-based models via introducing the basic message passing as well as model improvement methods including multi-head attention mechanism and residual structure. At last, we give task-oriented evaluation metrics for DL-enabled wireless resource allocation schemes. We also highlight certain challenges and potential research directions for the application of GNNs in wireless networks.
\end{abstract}

\section{Introduction}
The development of smart radio environment (SRE) enabler and high-frequency communication complicates the wireless network architectures and propagation of radio frequency signals. It becomes increasingly crucial and challenging to facilitate wireless resource allocation (including interference management and signal processing) to satisfy the quality of service (QoS) requirements of intelligent applications \cite{6G}. Traditional convex (CVX) optimization methods are computationally complex on account of their iterative frameworks and instance-by-instance property, which are hard to meet the time-intensive demand. In addition, when it comes to complicated wireless optimization, the mathematical approaches may rely on frequent relaxation and approximation and thus, mismatch the original problem. In the view of that the artificial intelligence (AI) achieves huge success in many application fields, AI empowered wireless networks has the potential to give birth to innovative and novel wireless optimization technique\cite{ai,new0}. Recently, some existing works have attempted to apply DL to solve the resource allocation problem in wireless communication\cite{dl}. Although some  classic neural networks such as multi-layer perceptrons (MLPs) and convolutional neural networks (CNNs) can be trained to realize near-optimal and real-time  optimization under learning-to-optimize paradigm, they may suffer from poor generalization performance in dynamical wireless networks. The problem-driven discipline of DL motivates the development of specialized neural network architectures for wireless networks. By taking the graphical topology of wireless networks into consideration, the graph neural networks (GNNs) tailored to operate directly on graph-structured data have drawn increasing attention \cite{GNN1,new1}.

There are good advantages of GNNs over other neural networks from the perspective of resource allocation in wireless networks. First, the permutation invariance and equivariance of GNNs simplify the input and improve the utilization efficiency of  samples such that the training cost is reduced \cite{new3}. Second, the node/edge-level readout operation and weight sharing of the GNNs enhance the generalization capability over the number of nodes/edges and the application to unseen problem sizes. Besides, the node/edge-level readout operation is the key factor to facilitate the parallel execution and the distributed implementation. Third, the GNN provides a paradigm to explore the underlying relationships by the interaction among elements in graphs which provides more degrees of freedom to utilize the prior knowledge of theoretical results in optimization of wireless networks. Forth, by regarding GNN as a feature extracting module, optimizing different system utilities can reuse the same extracting module under the unsupervised learning framework. Nevertheless, most existing works on GNN-enabled wireless optimization employs the vanilla GNN in wireless networks or problems. Some limitations restrict the development of GNN-based models in more complicated scenarios. Therefore, an in-depth study on GNNs for wireless networks is required.

There are some existing overviews on GNNs which focus on the application of the GNN in wireless networks\cite{GNN,GNN2}, while this article pays more attention to improve the capability of GNN-based models for optimizing wireless networks. To the best of our knowledge, little effort has been made on graph representation, model architectures and evaluations. The three aspects are fundamental to develop and improve the GNN-based models for wireless networks.
\begin{itemize}
\item The graph representation is to organize the wireless network parameters into graph-structured, which has a great impact on the scalability of the GNN-based models. Besides, the definition of node/edge features and the interaction over graph is determined by the graph representation.
\item The model architectures play the central role in developing GNN-based models to learn and exploit relational patterns between the different elements within wireless networks. Designing model architecture by taking the task-specific properties into account contributes to not only training efficiency but also expressive performance.
\item The evaluation for DL-based methods are different from the theoretical approach, as they cannot guarantee stationary-point solutions. Besides, the comprehensive evaluation is the basis to compare DL models, which helps to promote the DL-enabled resource allocation schemes.
\end{itemize}

\begin{figure*}[t]
\begin{center}
{\includegraphics[ width=.95\textwidth]{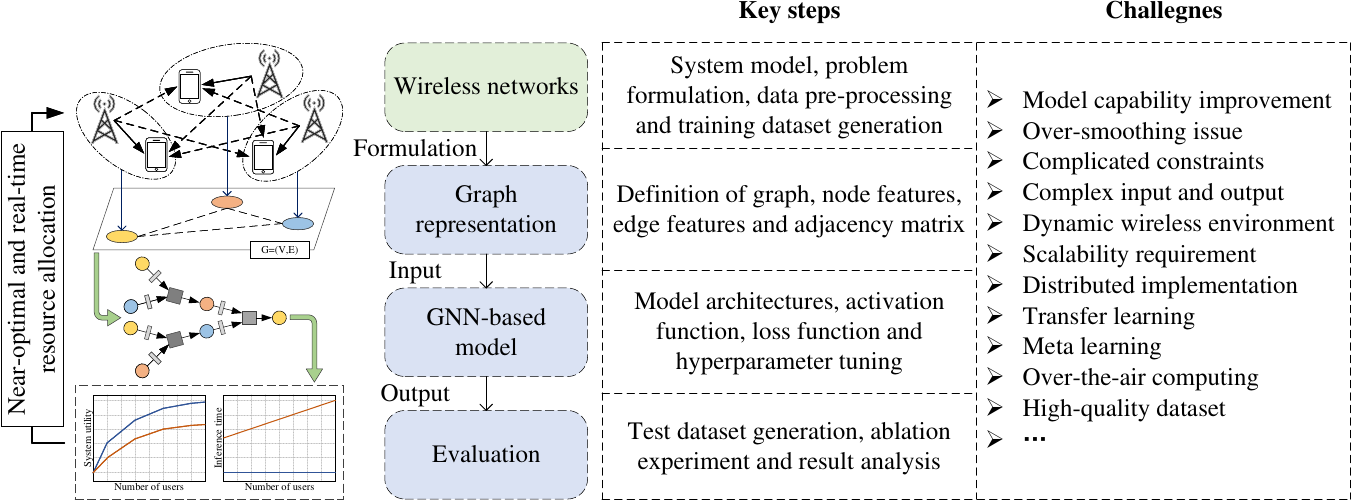}}
\caption{Framework of GNN-enabled wireless optimization.}
\label{process}
\end{center}
\end{figure*}

Based on the three aspects, the article intends to give a uniform framework to facilitate the application of GNN in wireless networks as shown in Figure \ref{process}. Besides, we list the challenges in the key steps of the framework. Some can be tackled well by this article while some require further investigation. For example, the model capacity of GNNs is required to be improved such that they can be used in various wireless networks. The over-smoothing issue is an intrinsic problem of GNN-based models to stack deep layers \cite{new5}, while the complicated constraints and complex input and output are the characteristic of wireless resource allocation problem. To adapt to dynamics and scalability of wireless environment is the advantage of GNN-based models, but to quantify this capability is not easy. For large-scale networks, the distributed implementation of the resource allocation scheme is crucial and the over-the-air computing is very suitable to be integrated into the message passing mechanism in GNN to facilitate the distributed implementation. Transfer learning and meta learning can improve the expressive performance as well as reducing training overhead. At last, the high-quality dataset is necessary as the DL-based method is data-driven.


\section{Graph Representation}

\begin{figure}[t]
\centering
\includegraphics[width=0.49\textwidth]{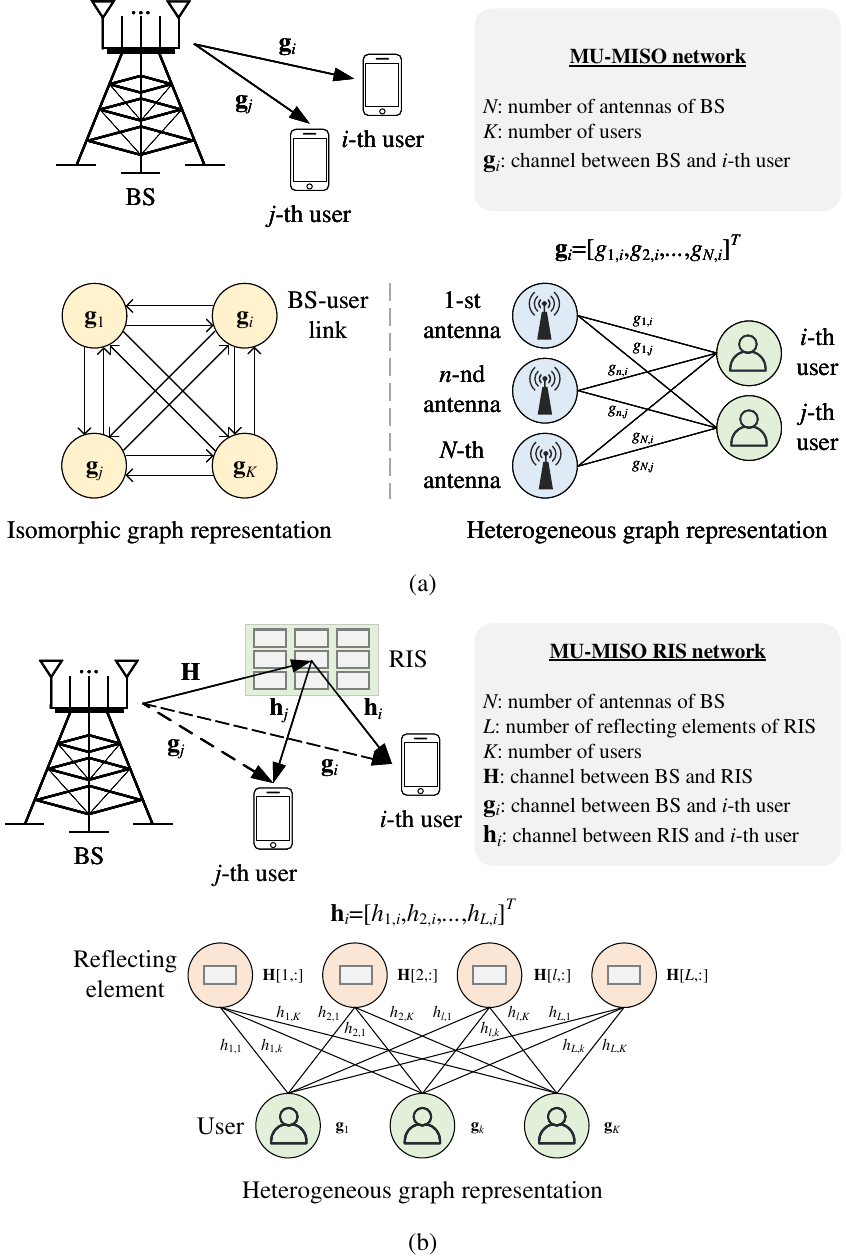}
\caption{(a) Isomorphic and heterogeneous graph representations of MU-MISO network. (b) Heterogeneous graph representation of MU-MISO RIS network.}
\label{rep}
\end{figure}

The wireless networks are usually with topological structure, which can be represented by graphs. Formally, a graph denoted by ${\cal G}=({\cal V},{\cal E})$ is defined by a set of notes ${\cal V}$ and a set of edges ${\cal E}$ between the nodes. In general, each node is with a node-level feature while the relationship between nodes is represented by the adjacency matrix and the edge features. Therefore, the graph representation is to model the elements (such as antenna, user, wireless link) in wireless networks as nodes and build the relationship between nodes to be edges. Note that different graph representations result in not only different node/edge features and adjacency matrices but also different interaction between nodes. Therefore, the graph representations of wireless networks play an important role in GNN-enabled wireless optimization.

Based on the types of nodes, we divide the existing graph representations of wireless networks into two categories, i.e., the isomorphic graph representation and the heterogeneous graph representation.

\subsection{Isomorphic graph representation}

In the isomorphic graph representation, there is only one type of nodes, which means that the wireless networks support one kind of services (such as information transmission, energy harvesting, radar sensing) or include one type of links.

The advantage of the isomorphic graph representation is that the GNN-based models accepting isomorphic graphs are efficient to train while an intuitive disadvantage is that it is hard to represent the complex wireless networks and scale to multiple elements. For instance, the multi-user multi-input-multi-output (MU-MISO) system can be represented by a fully connected isomorphic graph as shown in Fig. \ref{rep}(a) where each node represents the link between the base station (BS) and one user with the node feature of the corresponding channel state information (CSI) and each edge represents the existence of the inter-link interference. Note the edge features can be null or with manually setting value representing the inter-link relationship.

\subsection{Heterogeneous graph representation}

In the heterogeneous graph representation, nodes are imbued with types, and edges usually connect nodes of certain types. As a result, the heterogeneous graph is capability to represent various types of elements and services.

The advantage of the heterogeneous graph representation is that it can represent complicated wireless networks and multiple relationships while the disadvantage is that it challenges the architecture design and training stability. Besides, the GNN-based models are usually with the scalability to nodes. Therefore, the GNN-based models accepting input of heterogeneous graphs are expected to be with more powerful scalable capability, such as users, antennas, reflecting elements. Note that the  MU-MISO system can also be represented by a bipartite unweighted graph (i.e., a heterogeneous graph) as shown in Fig. \ref{rep}(a), where there are two types of nodes, i.e., antennas and users, and the edges connect nodes with different types. Another example of heterogeneous graph representations is for the MU-MISO reconfigurable intelligent surface (RIS) network as shown in Fig. \ref{rep}(b), where the reflecting elements and the users are modeled as nodes with different types.

With the definition of graph representation, the resource allocation problem can be reformulated into graph optimization problem which can be solved by GNN-based models. Moreover, instead of using the vanilla system parameters, some pre-processing \cite{feature} or feature enhancement method can be utilized to re-construct the node/edge features \cite{feature2}.

\section{Architecture of GNN-Based Models}

\begin{figure*}[ht]
\begin{center}
 {\includegraphics[ width=0.98\textwidth]{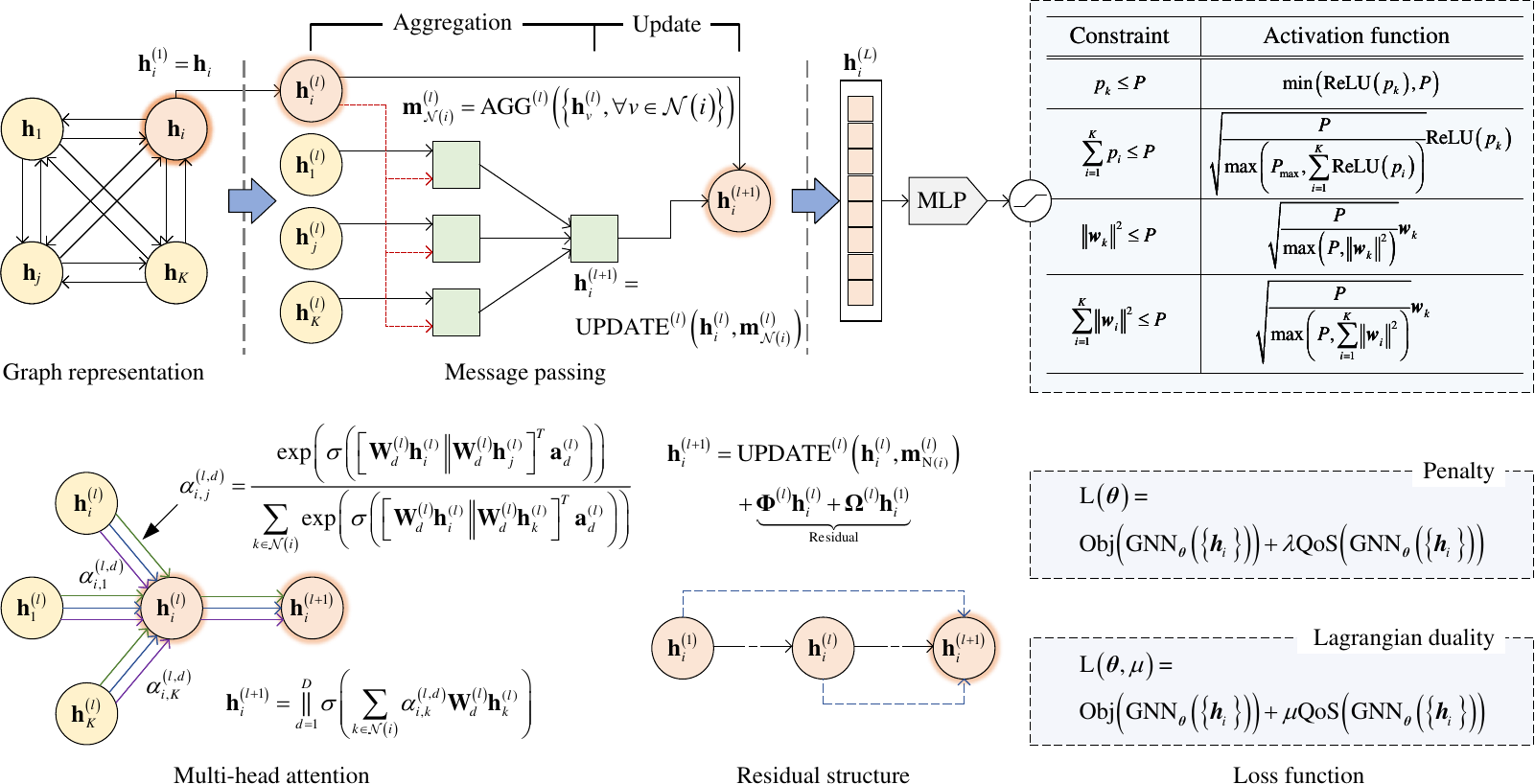}}
\caption{Architecture of GNN-based models: message passing, multi-head attention mechanism, residual structure, AF and loss function \cite{GRL}.}
\label{GNN}
\end{center}
\end{figure*}

With the graph representation, a GNN-based model{\footnote{The resource allocation problem can be tackled by  Node-GNN for node-level task and Edge-GNN for edge-level task\cite{vertexedge}. As most edge-level tasks can be equivalently transferred into node-level tasks, this article focuses on the architecture design of Node-GNN.}} is to map the graph-structured network parameters to the desirable resource allocation scheme. The most popular GNN-based model is message passing neural network (MPNN) or graph convolutional network (GCN). However, these vanilla GNNs face difficulties of inflexible relationship between nodes and stacking deep layers. To improve the model capability of GNN-based models, some mechanisms such as attention and residual can be taken into account for designing the model architecture. Besides, different from the applications of DL in other fields, one of the challenges of DL-enabled wireless optimization is that the output of the neural networks are required to satisfy the constraints (such as power budget, QoS requirements, etc). To guarantee feasible output, the specialized activation functions (AFs) or/and loss functions are required.

To describe the detailed processes including massage passing, attention and residual, we refer to an illustrative example as shown in Figure \ref{GNN} where the input graph of the GNN is the isomorphic graph representation of the MU-MISO network.

\subsection{Message passing (vanilla GNN)}

The message passing is a key component to build up GNN-based models. Particularly, the message passing involves two steps: aggregation and updating. During each message passing iteration, the (hidden) feature corresponding to each node is updated by combining the message aggregated from its immediate neighbors and its previous feature (via self-loop). As the message passing iterations progress, each node intends to contain more and more information from further reaches of the graph, which result in more useful node features. With the learned node features, the solutions can be obtained via readout operation which can be realized by a decoder, like MLP.

To improve the GNN performance, the main idea is to propose novel architectures of aggregation and updating.

\subsection{Multi-head attention}

The multi-head attention mechanism is a widely used strategy to enhance the aggregation \cite{gat}, especially in cases where some neighbors may be more informative than others.

The traditional message passing aggregates the message from node neighborhoods with equal weights which may not be able to reflect the different influences among nodes. Specially, in wireless networks, one user receive different interference from different users. The inter-user interference is implicit but hidden within the correlation of channel state information of users. The  attention mechanism is to assign an attention weight or importance to each neighbor, which is used to weigh this neighbor's influence during the aggregation step.

By the multi-head attention mechanism, i.e., using multiple independent attentions and concatenating the results, the model capacity and stability can be further improved.

\subsection{Residual structure}

The residual structure is an effective strategy to enhance the updating, especially in cases where nodes all fully connected.

To extract high-level node representations, multiple massage passing iterations are required to stacked, which however, induces the over-smoothing issue to degrade the effectiveness of the GNN-based models \cite{oversmooth}. The essential idea of over-smoothing is that after several iterations of message passing, the representations for all the nodes in the graph can become very similar to one another. To alleviate the over-smoothing issue in GNNs, the residual structure allows the updating process to preserve some information from previous iterations of message passing such that the node-level features keep distinguishable.

\subsection{Activation function and loss function}

Although the output of GNN-based models can be shaped into the required dimension via MLP, it may not satisfy the constraints. One method to yield feasible output is to design AFs to project the infeasible output to be feasible. Figure \ref{GNN} gives some parameter-free  AFs to handle the constraints on power budget. However, for some complicated constraints, it may be hard to guarantee the feasible output of the DL models by activation. In this cases, one strategy is to embed these constraints into the objective function to construct the loss function via penalty method (PM) or Lagrangian duality  method (LDM). By adjusting the penalty hyperparameter or learning the Lagrangian multipliers, the output of the DL models are very likely to feasible.

The DL models for wireless optimization can be trained via unsupervised learning, which not only reduces the training cost but also allows sharing of the training set for different system utilities. Besides, the different system utilities can also be transfer learned between each other.

A well-trained GNN-based model can yield a resource allocation result for its input  via real-time computations. However, the resource allocation result may be with good or poor performance or even infeasible. Therefore, a compressive evaluation for the GNN-based models is required.

\section{Evaluation Metrics}

The main goal of the DL-based methods for wireless networks is to realize near-optimal and real-time optimization. Therefore, the evaluation of the  DL-based methods is of high importance. However, the existing evaluation metrics, such as convergence behavior, optimality and computational complexity, are mainly used to evaluate the CVX-based algorithms and may not be enough to evaluate the DL-based methods. On the other hand, as a widely used methodology, the DL has its specific evaluation metrics, such as ablation experiment and generalization performance. Therefore, we intend to develop some new metrics to comprehensively evaluate the DL-based wireless optimization methods by taking both wireless communications and DL into account.

\begin{figure}[t]
\begin{center}
{\includegraphics[ width=.5\textwidth]{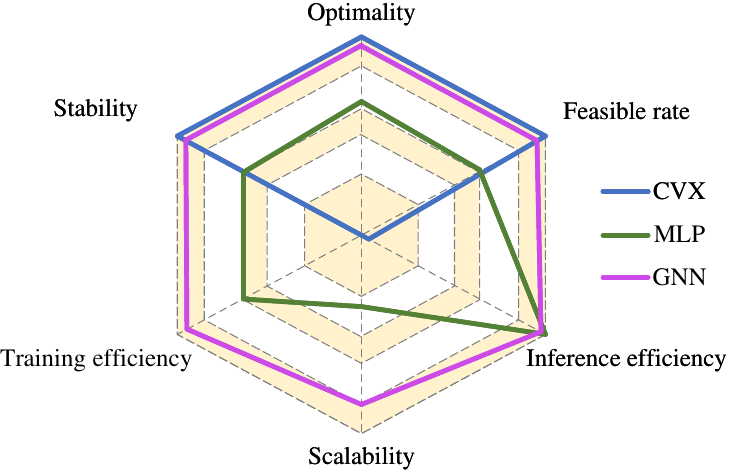}}
\caption{Multi-dimension evaluation metrics.}
\label{eva}
\end{center}
\end{figure}

\begin{itemize}
  \item [1)]
  \emph{Optimality:} The ratio of the average objective value (with feasible solutions) by the DL-based method to the average optimal objective value (by the CVX-based methods or exhaustive search methods) over the test samples. Actually, the optimality performance is equivalent to the generalization performance when the training set and the test set have the identical setting, and the optimal objective value can be regarded as the label of the test sample. The demand on the feasible solutions is necessary since the infeasible solution shall result in better objective value and induce unfair comparison. As the DL-based methods cannot always yield global optimal solutions, the ratio should be lower than $100\%$. However, for some complicated problems, the optimal objective value may be unavailable for all test samples. In this case, the results by CVX-based methods or existing schemes can be approximately utilized as labels and the ratio may exceed $100\%$.
  \item [2)]
  \emph{Feasibility rate:} The percentage of the feasible solutions to the considered problem by the DL-based methods over the test samples. The feasibility rate is an important metric especially when the optimization problems involving complicated constraints which are not handled by activation. A DL-based method  with high optimality performance but low feasibility rate is unacceptable. Besides, the value of the QoS threshold has a great impact on the feasibility rate during the testing.
  \item [3)]
 \emph{Inference efficiency:} The average running time required to calculate the feasible solution for the considered problem by the DL-based methods. Generally, the well-trained DL models are with much faster inference speed than the CVX-based methods. Nevertheless, the wireless networks require real-time resource allocation, which makes lightweighting  DL models a central consideration. Besides, a trade-off exists between inference efficiency and model capability.
 \item [4)]
  \emph{Scalability:} Evaluate the optimality performance of the DL-based methods on the test set with different settings from the training set, such as different numbers of users and different values of power budget. The scalability is a key requirement in the dynamic wireless networks which is also the advantage of the GNN-based models.
  \item [5)]
  \emph{Training efficiency:} The training samples and the  epochs required to coverage. Most DL models are data-driven. Although the unsupervised learning alleviates the demand on labels of the training samples, it still requires unlabeled samples to train the models. Via model architecture design and prior knowledge, the training efficiency  can be enhanced.
  \item [6)]
  \emph{Stability:} The percentage of the test samples (with feasible solutions by the DL methods), which suffer from $\le n\%$ ($n\in(0,100)$) performance loss to the corresponding labels, in all test samples with feasible solutions. The optimality reflects the average performance while the stability reflects the variance performance.  A poor stability performance may be caused by unsuitable dataset setting or hyperparameter selection.
\end{itemize}

An example of the multi-dimension evaluation metrics is illustrated in Fig. \ref{eva} to  compared different DL models and CVX-based method, from which it is observed that it is hard to find a model that outperforms other models in terms of all metrics. Therefore, the model should be selected based on the requirements of specific application. Namely, the DL-enabled wireless optimization is to develop the ``best-effort" DL models.

\section{Case Study and Numerical Results}

This section provides numerical results by a case study on MU-MISO networks to verify the effectiveness of the message passing, attention-assisted aggregation and residual-assisted updating as well as validating the scalability of the GNN-based models. Three classic system utility optimization tasks are considered i.e., sum rate maximization (SRM), energy efficiency maximization (EEM) and max-min rate (MMR). Besides, the AF method, PM and LDM for handling sum-power constraint are compared, and the supervised learning and unsupervised learning are also compared.

\begin{figure*}[ht]
\begin{center}
 {\includegraphics[ width=1\textwidth]{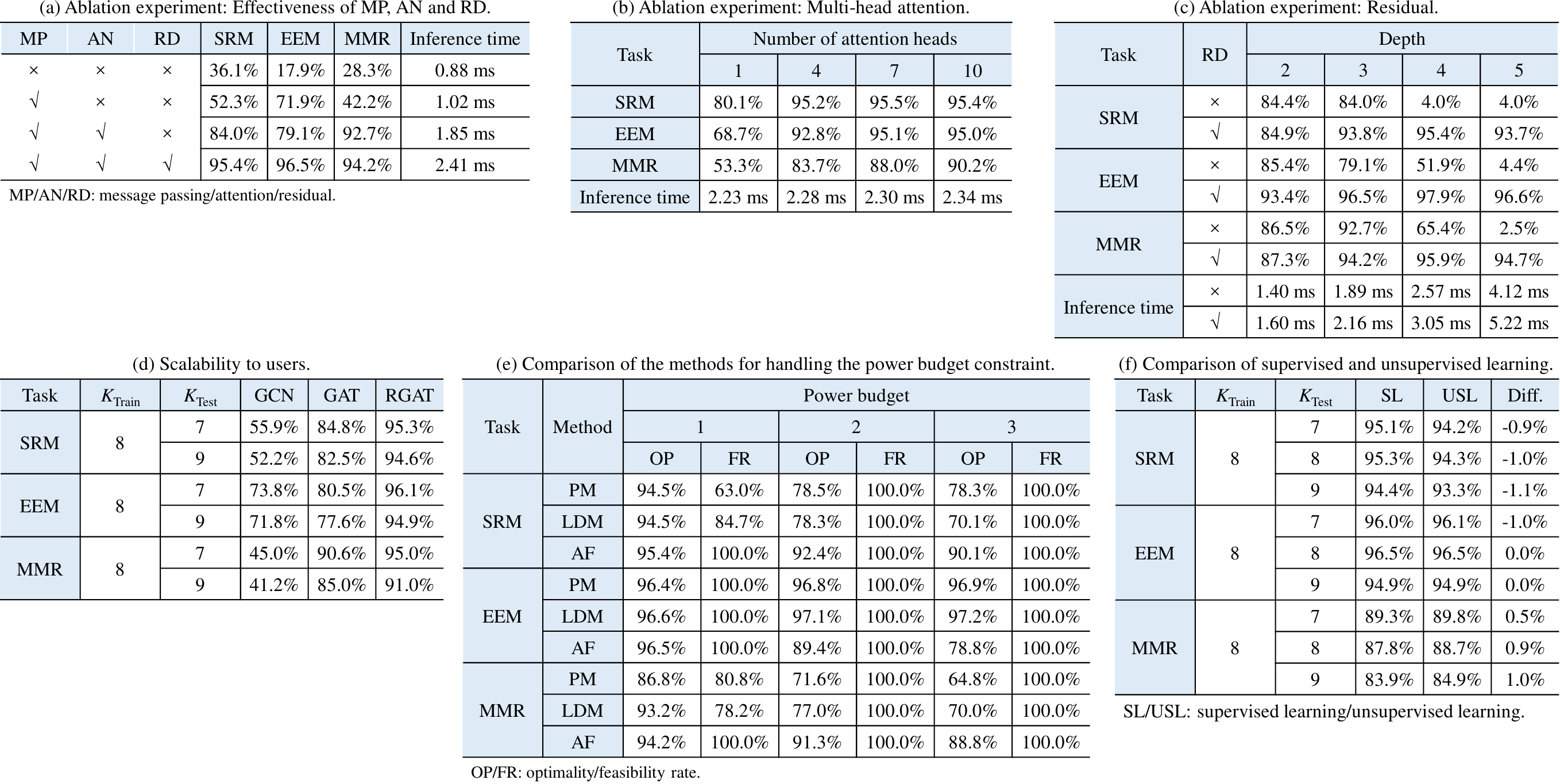}}
\caption{Numerical results.}
\label{nr}
\end{center}
\end{figure*}

Figure \ref{nr}(a) provides an ablation experiment to show the effectiveness of the message passing, attention and residual in an $16$-antenna $8$-user MISO network in terms of the three system utilities. It is observed that the message passing, attention and residual could improve the optimality of the models, especially for MMR, which validates the feature extracting capability of the GNN-based models over wireless networks. Due to involving more computation processes, the message passing, attention and residual also slightly increase the inference time. Furthermore, the ablation experiment on multi-head attention mechanism and the residual are respectively presented in Figure \ref{nr}(b) and Figure \ref{nr}(c). It is observed  that with increase of number of attention heads, the optimality performance is improved while the performance gain becomes smaller. Besides, the optimality performance is improved with depths with the assistance of residual while degraded with depths without the residual. 

The scalability is the key advantage of the GNN-based models over other DL models in the dynamic wireless networks. Figure \ref{nr}(d) shows the scalability performance to users of the three GNN-based models, i.e., GCN, GAT \cite{li_gat}, RGAT\cite{li_GNN}, for the optimization of  considered system utilities. All GNN-based models are trained with $8$ users while tested with $7$ and $9$ users. It is observed that all GNN-based models are able to generalize to unseen problem sizes in the training set. The reason is that the trainable weights of the GNN-based models can be independent of the graph size. Moreover, with sophisticated  architecture design, the GAT and RGAT achieve better scalability performance.

Figure \ref{nr}(e) compares three methods, i.e., AF method, PM and LDM, to make the output of the DL models satisfy the sum-power constraint for beamforming vectors. It is observed that the AF method always guarantees $100\%$ feasibility rate while the PM and LDM may fail to yield feasible solution for some realizations. Besides, in most cases, the increment of the power budget induces optimiality loss as the enlarged feasible set complicates the learning task, and the AF method slightly outperforms the other methods in the term of the optimiality. Nevertheless, the AF method may be hard to implement for complex constraints which however, can be handled by the PM and LDM.

Figure \ref{nr}(f) shows that the supervised learning and the unsupervised learning achieve very close  optimality and scalability performance fot the tested DL models. The reason is that the DL model the utilize gradient decent method (like the CVX-based method) to lower the loss function. To minimize the difference between the objective value and the optimal value (via supervised learning) is equivalent to minimize/maximize the objective function (via unsupervised learning). The unsupervised learning  not only simplifies the training set (compared with supervised learning) but also prevent complicated problem reformulation (compared with deep reinforcement learning), which makes the focus of the DL-enabled optimization in wireless networks to be developing more powerful neural networks.

\section{Conclusion and Future Directions}

In this article, we provided a comprehensive overview of applying GNNs to optimize wireless networks and discussed its key steps. In particular, the graph representation of the wireless networks, the principal and design frameworks, as well as evaluation metrics are detailed presented.  Numerical results demonstrated the advantages of the GNN-based models in comparison with CVX optimization method and other DL models.

Though GNN-based wireless optimization showed significant performance advantages in providing near-optimal performance, scalability to network typologies and real-time inference, there are still some key issues which need to be addressed as described in the following.

\textbf{Graph representation of complicated wireless networks:} Although there is a trend on applying the GNN to wireless networks, most of existing works consider the classic application scenarios. For multi-hop, multi-service and multi-element networks, the graph representation still remains an open issue. Besides, instead of directly using the system parameters as node/edge features, how to manually design the node/edge features based on the prior knowledge is also an important question to investigate in graph representation.

\textbf{Efficient approach to handle complicated constraints:} Yielding feasible solutions challenges the DL-enabled resource allocation schemes, especially when there are multiple complicated constraints and the output involves complex values, e.g., beamforming vectors and phase shift elements. For GNN-based models, the feasible rate may be degraded when they face unseen problem sizes. One promising way is to reformulate the constrained optimization problem into multi-objective problem and leverage the multi-task learning to enhance the robustness of GNN-based models. Besides, the transfer learning is also a choice to improve the feasible rate at little fine-tuning cost.

\textbf{Powerful GNN-based structure for ultra-dense networks:} Most of existing GNN-based models are designed for the single-cell or D2D scenarios, while these models are hard to extend to the practical ultra-dense networks. Some new mechanisms (like \cite{new2,new4}) are required to develop more powerful GNN-based models. Then, the trade-off between inference performance and inference speed should be the central consideration for designing the architectures of GNN-based models. Another promising way is to employ the over-the-air computing framework to enable the distributed implementation of the GNN-based models. Besides, the message passing mechanism requires to re-deigned by taking the signaling overhead exchanging process into account.

\textbf{Engineering-awareness dataset and evaluation metrics:} Different from the CVX-based methods which aims for theoretical results, the DL-based methods are responsible to meet the practical implementation requirements. Training DL models often requires huge amounts of samples with enough diversity to abstract deep insights, which makes the engineering-awareness dataset necessary. Besides, the evaluation metrics in terms of practical implementation play the key role to connect the theoretical research and the practical demands. The standard datasets and evaluation metrics  are also helpful to build up the open source DL libraries and platforms for wireless networks.

\textbf{GNN-enabled deep reinforcement learning:} Deep reinforcement learning (DRL) is an important branch of DL, which has been widely adopted to solve the complicated resource allocation problems in RIS- and UAV-assisted networks. Different from unsupervised learning, the DRL enables the online training of the model. Most existing DRL methods train MLPs to generate state value or action. As the GNN-based models have been proved to be with more powerful expressive capability, training GNN via DRL shall be a hot research spot. Note that the inputs, i.e., graph representations, of GNN-based modes trained by the unsupervised learning and the DRL may be different even for the same problem. Besides, the comparison between DRL-based GNN and unsupervised learning based GNN remains an open issue.

\end{document}